# Exploration of Recent Advances in the Field of Brain Computer Interface

M.Rajyalakshmi[1], T. Kameswara Rao[2] and Dr. T. V. Prasad[3]

[1&2]*Assoc. Prof., Dept of Comp. Sc. & Engg., Visvodaya Technical Academy, Kavali, AP, India*
[3]*Dean of Computing Sciences, Visvodaya Technical Academy, Kavali, AP, India*
Email: rajyalakshmiphd@gmail.com, tkrphd@gmail.com, tvprasad2002@yahoo.com

***Abstract** – A new approach for implementing number of expressions, emotions and, actions to operate objects through the thoughts of brain using a Non-Invasive Brain Computing Interface (BCI) technique has been proposed. In this paper a survey on brain and its operations are presented. The steps involved in the brain signal processing are discussed. The current systems are able to present few expressions and emotions on a single device. The proposed system provides the extended number of expressions on multiple numbers of objects.*

## 1. WORKING OF BRAIN

The part which set humans apart from all the living species by allowing us to do wonders and to achieve great things is the Human Brain. Brain is one of the most important parts of the body that controls entire actions of the human. Brain is responsible for the tasks made by human. Activities done by brain include from a complex task heart beat rate, emotions, learning and control etc. Ultimately, it is the one which makes us human and which shapes our thoughts, innovations and ideas. [1]

Brain is composed of four major parts viz., cerebrum, cerebellum, cerebral cortex and the brain stem. Figure 1 shows the four major parts of brain. On each hemisphere of cerebrum it has four lobes, which are called as frontal lobe, parietal lobe, temporal lobe and occipital lobe. The brain stem has two divisions, the pons and medulla. [2]

- Frontal lobe is responsible for many functions like reasoning, planning and problem-solving. Basic movements are also controlled by the frontal lobe. It also responds to emotional situations. It is also responsible for consciousness, activities related with the environment and personality.
- Parietal lobe reacts to information obtained from touch, pressure, temperature and pain. It is also responsible for the visual attentions and moving voluntarily.
- Temporal lobe is the language comprehension center. Its functions include smell, hearing and categorizing objects.
- Occipital lobe is responsible for the vision, identification of color and movement of an object.

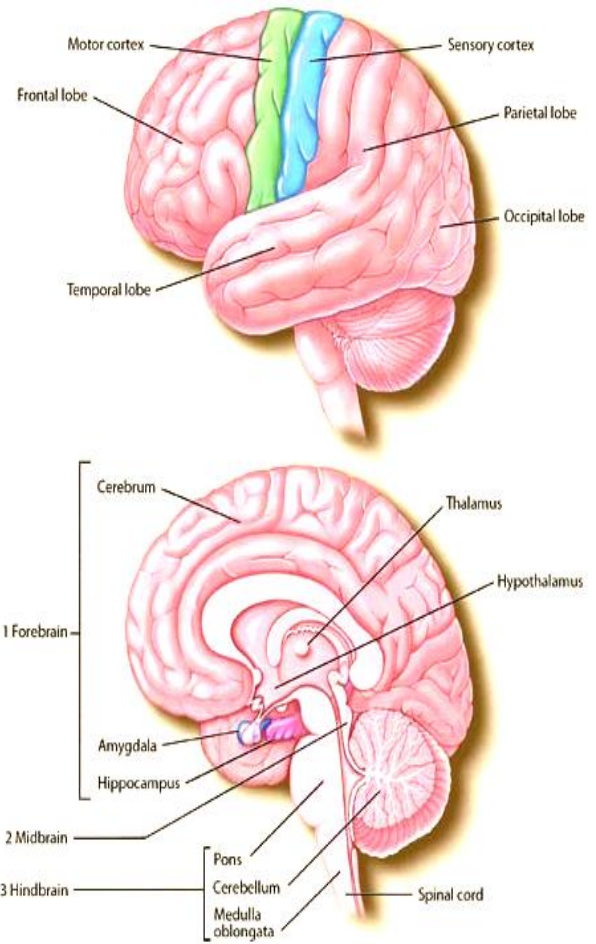

Figure 1 Four major parts of the brain

The balance, posture and equilibrium of human body are controlled by cerebellum. Pons is section of brain stem which is responsible for movement in eyes and face. Medulla controls the most important functions such as heart rate and breathing. [3]

## 2. BCI: DEFINITION AND WORKING

The advancement of technology has brought a new reality. Today, the technology that allows humans to use the electrical signals to record the brain activity to interact with, influence, or change their environments is known as Brain Computer Interface (BCI) [4], which acquires the brain signals, performs analysis, translation, and classification of the signals to do the desired action. BCI systems are of two types: Invasive and Non-Invasive. [5]

Invasive BCI systems interact with brain directly by penetrating electrodes into the brain cortex. This requires surgery and the results have to be obtained with in stipulated time period. These are used in US and they are expensive.

In Non Invasive BCI systems, the electrodes are placed on the scalp and the brain activity is recorded by using Electroencephalography (EEG). It is less expensive than Invasive BCI system. Non invasive BCI systems are primarily evolving from European and Asian efforts.

The goal of BCI is to enable people with neural abilities that have been damaged by amputation, trauma and disorders to function properly through prosthetic limbs or robotic devices. [6]

The brain signal recording in invasive BCI can be done by using action potentials from nerve cells or nerve fibers, synaptic and extracellular field potentials, electro corticograms (ECoG). The brain signal recording in non invasive BCI can be done by using electroencephalography (EEG), functional Magnetic Resonance Imaging (fMRI), magnetic Sensor Systems, slow cortical potentials (SCP), Thermo-graphy, near Infra Red Spectrum (NIRS). Various types of electrodes used to record brain signals are shown in Figure 2. In figure 2, A, B, C show the EEG electrodes that can fitted into EEG caps, D shows an ECG/EMG electrode which is placed close to the muscle or heart. Electrodes of type A and D can also be used for EOG recordings. [7][8]

Non invasive BCI systems that use EEG signals to record activity are of mainly two types: P300 systems and sensory motor rhythm system.

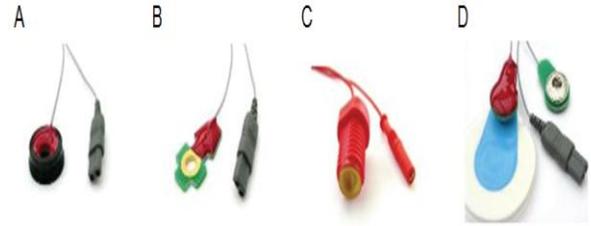

Figure: 2 Electrodes for EEG, ECG, EOG, measurements. A: Active single electrode with multi-pole connector; B: active gold electrode with multi-pole connector; C: screw-able passive gold electrode to adjust location; D: active ECG electrode with disposable Ag/AgCl electrode

In P300 systems a wave of voltage occurs within 300ms of a perceived event. P300 systems will be affected by the external factors such as fatigue. EEG recordings are converted to virtual key board signals using this P300 system.

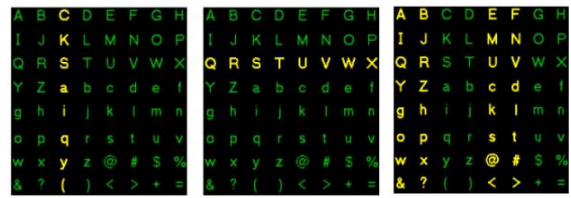

Figure 3: P300 run time interface

In sensory motor systems, the cursor movements are controlled by the brain. The electrodes are attached to the scalp above the sensory motor cortex, part of the brain region that is responsible for perceiving information and for performing voluntary movements. [5]

A BCI is system actually consisting of a sensor, a natural detector or translator, and an actuator to carry out an action. Sensor is dedicated to observe, either indirectly or directly, changes in neural activity related to the intent to influence an external device, the actuator. Actuator is a type of motor for moving a mechanism or system. BCI include a cursor on a computer screen, a motorized wheel chair, a semi autonomous robot, a prosthetic limb or functional electrical simulation devices that could reanimate a paralyzed limb. Between sensor and actuator lies the decoder that receives neural data recorded by sensor and discerns user intention and converts that into a command signal for the actuator. The choice of sensor, decoder and actuator are interdependent in the designing of a BCI system. BCI devices detect and translate neural activity into command sequences for computers and prostheses. The aim of these devices to restore function in patients suffering from loss of motor control i.e., stroke, spinal cord injury, multiple sclerosis and amyotrophic lateral sclerosis (ALS).[9]

BCI as a communication system allows the individual to send messages or commands to the external world do not pass through the brain's normal output pathways of peripheral nerves and muscles. For example, in an EEG-based BCI the messages are encoded in EEG activity.

A BCI system that does not use the brain's normal output pathways to carry the message, but activity in these pathways is needed to generate the brain activity (e.g. EEG) that does carry it is the dependent BCI. A BCI system that does not depend on the brain's normal output pathways is an independent BCI. The message is not carried by peripheral nerves and muscles, and, further-more, activity in these pathways is not needed to generate the brain activity that does carry the message. [15]

In non invasive BCI systems, the EEG signals can be detected only when many of neurons are synchronized. An EEG signal is the surface potential variation that is recorded by the electrodes placed on the scalp.

Analyzing the brain waves or EEG signals is very complex because it includes large amount of information received electrodes. Although there is more amount of information, the brain waves are categorized into the following types based on their emanations. Among these waves five of them are very important. They are:

- **Beta:** Beta waves frequency is ranging from 13 to 30Hz and has a low voltage between 5-30μV. Beta wave is generated during active thinking, active attention and, solving problems. It can reach 50Hz frequency in the intense mental activity.

- **Alpha:** Rate of change of frequency of these alpha waves is in between 8 to 13 Hz with a voltage of 30-40μV. Relaxedness and inattention are indicated by these waves. Alpha seems to be an empty mindless state that can be eliminated by opening eyes, by hearing unfamiliar sounds.

- **Theta:** The frequency of Theta waves is ranging from 4 to 7 Hz, with amplitude greater than 20μV. It is associated with frustration, unconsciousness, disappointment and deep meditation.

- **Delta:** Delta waves are emitted within the range of frequency 0.5 to 4Hz and with variable amplitude. Deep sleep, walking, defects in brain are indicated by these waves.

- **Gamma:** Gamma waves are emitted within the range of 35Hz and above. These waves reflect the consciousness.

- **Mu:** These waves are in range same as that of alpha waves. It is an EEG activity associated with the motor activities that are recorder on motor cortex. These are recorded on Occipital lobe of the brain. [10]

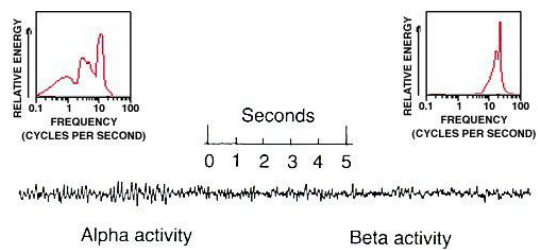

Figure 4: Alpha and Beta waves activity generated during the EEG recording of brain signals.

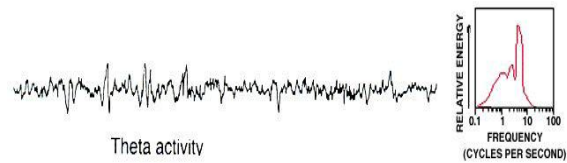

Figure 5: Theta waves during the EEG recording of brain signals.

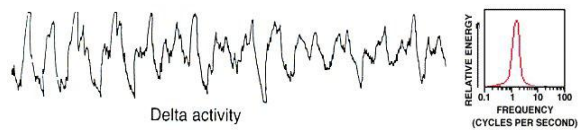

Figure 6: Delta Waves generated during the EEG recording of brain signals.

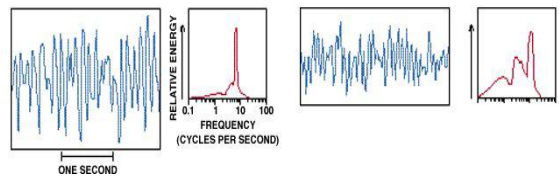

Figure 7: Gamma and Mu waves during the EEG recording of brain signals.

## 3. ARCHITECTURE OF BCI

Architecture of a BCI system is majorly composed of two mechanisms: brain signal acquisition, brain signal pre-processing. Brain signal processing again includes two mechanisms: feature extraction and translation algorithm.

a) **Brain Signal Acquisition:** The signals acquired from invasive and non-invasive methods are amplified and then sampled. The first layer of the BCI architecture is the brain signal acquisition. It is responsible for capturing EEG signals from the subject. The current implementations using Emotiv EPOC neuro-head set to acquire these signals. The Emotiv EPOC neuro-head set is a wearable EEG device which collects signals using 14 electrodes. The information is then transmitted wirelessly to a PC over Bluetooth (IEEE 802.15.1). [12] [14]

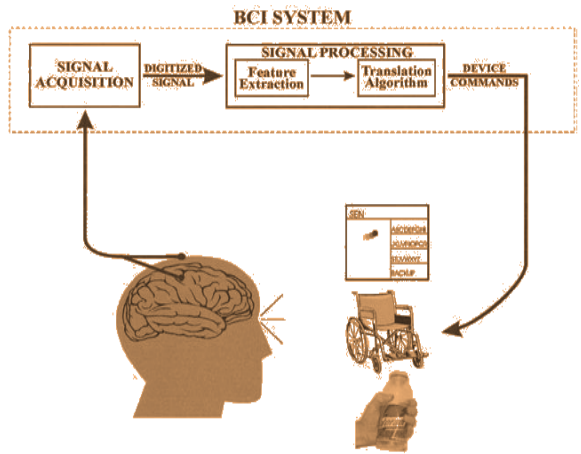

Figure 8: Architecture of BCI

In the last few years, several EEG-based gaming devices have made their way onto the market and became available to the general public. Companies such as Emotiv Systems [22] and NeuroSky [23] are offering low-cost EEG-based BCI devices (e.g., Figure 9) and software development kits to support the expansion of tools and games available from their application markets.

Currently, there are more than 100 available applications ranging from accessibility tools, such as a mind controlled keyboard and mouse and hands-free arcade games, to so-called serious games, i.e., games with a purpose other than pure entertainment, such as attention and memory training games.

The novel EEG acquisition system in combination with an eye tracking system for hybrid operation is shown in Figure 9. This prototype consists of fixed holders for 22 electrode positions used for both SSVEP and, ERD/ERS, and two cameras used for eye tracking. Currently, EEG data are recorded using Ag/Ag-Cl electrodes and conventional electrolytic gel, and then digitized through a Porti32 amplifier. One of the future improvements of the system will be the EEG acquisition using water-based electrodes, which will make both daily setup and clean up much faster, easier and comfortable. In general, the main advantages of the acquisition system are the reduction of the subject's EEG preparation time, better repeatability of the electrode locations between EEG acquisition sessions, and the combination with eye tracking for hybrid control. The eye tracking system is used to detect the user's intentions (searching and selection) based on the viewing direction. If the subject is interested in a specific device in the environment, i.e., a predefined fixation time is reached, then the BCI system is used to interact with the selected device. [16]

b) **EEG Signal Pre Processing:** The need for signal pre processing is that the EEG recordings typically not only contain electrical signals from the brain, but also several unwanted signals [18-20], [24]:
- Interference from electronic equipment, as for example the 50 or 60Hz power supply signals.
- Electro myographic (EMG) signals evoked by muscular activity,
- Ocular artifacts, due to eye movement or blinking.

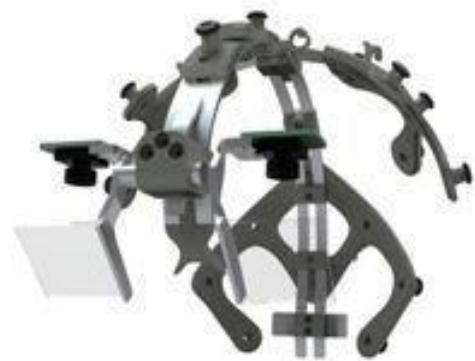

Figure 9: EEG signal acquisition system combined with eye tracking.

Those unwanted components may bias the analysis of the EEG, and may lead to wrong conclusions [19], [20]. Brain signal pre processing can be done by using three methods. They are Basic Filtering, Adaptive Filtering and Blind Source Separation. [21]

The signal pre processing block is fundamental to BCI when it is viewed as pattern-recognition system. Feature extraction is the part of brain signal pre-processing. In feature extraction, the similar objects are classified into the same class, and two different classes can be according to their differences. To identify the class of an object, it is important to extract some properties that reflect similarities as well as differences.

The signal processing module translates ongoing brain signals from 21 electrodes into device control commands. Eight signals are acquired from the visual cortex to detect SSVEP patterns using adaptive spatial filtering and the signal-to-noise ratio at each stimulation frequency. They are Pz, PO3, PO4, PO7, PO8, Oz, O9, O10. The remaining electrodes are used to detect ERD/ERS patterns during three types of motor imagery (MI): right hand, left hand, and feet imagery of movement. They include FC3, FCz, FC4, C5, C3, C1, Cz, C2, C4, C6, CP3, CPz, CP4, and AFz for a ground electrode The algorithm is based on auto regression, multiclass common spatial patterns and mutual information. [16]

c) **EEG Signal Classification:** The classifier is used to identify the intentions of the subject from a finite number of predefined choices. After the feature extraction a suitable classifier has to be designed. For performing classification EEG signals, Linear Discriminate Analysis (LDA) classifiers, Bayesian statistical classifier and, Fuzzy logic classifiers are used. [13]

## 4. BCI APPLICATIONS

BCI can be used in various areas like neuroscience, neurology, neuro engineering and, in various areas of medical sciences. BCI is relatively a novel technology with a potential to restore, substitute, or augment lost motor behaviors in patients with neurological injuries such as stroke, spinal cord injury, and traumatic brain injury, can cause chronic gait function impairment due to foot-drop. BCI can be mounted on wheel-chair devices, prosthetic limbs, mobile controlled robots, computer mouse control, and communication systems and, in environment control. [8][25]

Some of the BCI-like systems are lie-detection, alertness monitoring, neuro feedback, eye tracking, neuro marketing [26].

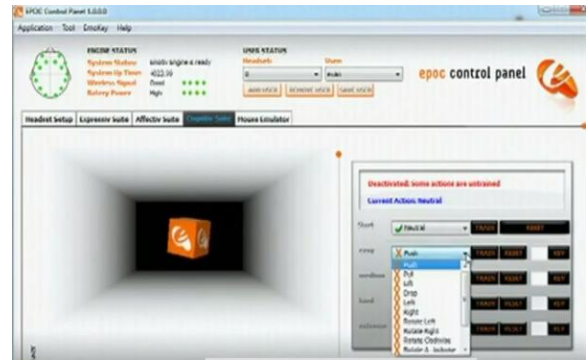

Figure 10: EPOC Control panel implemented by Tan Le [27]

**EPOC – The Mind Controlled Device:** Recently, Emotiv systems have introduced a mind controlled device with EPOC wireless head set. Figure 10 shows the interface how EPOC headset is used to operate devices through the thoughts of brain. This interface was implemented by Tan Le. In this interface the operations are performed on a single device. The operations that are performed on device are pull, push, show, hide, move left, move right, rotate and so on. The intention of user has to be maintained for duration of 8 seconds to observe the operation on the device. [27]

## 5. CONCLUSION AND FUTURE SCOPE

It is possible to use the BCI systems can be used in various applications areas and also it is capable of controlling devices through thoughts of brain. Current systems are able to operate single device with a few operations.

In future, devices that can control the working of home appliances which are operated by thoughts of human brain may be developed. Robots that will operate according to the emotions of the human may be developed.

## REFERENCES

[1] "Brain Facts, A Primer on the brain and nervous systems", http://www.sfn.org/skins/main/pdf/brainfacts/2008/brain_facts.pdf, 2008.
[2] "Brain-MindMachinery", available at www.worldscientific.com/ worldscibooks/10.1142/6704

[3] Lia Stannard, "Parts of the brain and their functions", Available: http://www.livestrong.com/article/77671-parts-brain-functions/, 2011.
[4] Jerry J.Shih;Dean. J. Krusienski: and Jonathan R, Wolpaw "Brain Computing Interfaces in Medicine", Mayo Clinical Proceedings, Volume 87, pp 268-279, 2012.
[5] Erica Westly, "Fixing the Brain Computer Interface", IEEE Spectrum, July 2010.
[6] Theodore W. Berger, John K. Chapin, Greg Patrick A. Tresco, "Brain-Computer Interfaces-An International Assessment of Research & Development Trends", Springer, 2008.
[7] Reza Fazel Rezai, "Recent advances in Brain Computer-Interface systems", InTech Publication, 2011.
[8] Niels Birbaumer, "Breaking the silence: Brain Computer Interfaces (BCI) for communication and motor control", Psychophysiology, 43, 517–532, 2006.
[9] Leigh R. Hochberg, John P. Donoghue, "Sensors for Brain Computer-Interfaces", IEEE Engineering and Medicine, 2006.
[10] Jorge Baztarrica Ochoa, Gary Garcia Molina, Touradj Ebrahimi, "EEG signal classification for Brain Computer Interface Applications", 2002.
[11] Sumit Soman, Soumya Sen Gupta, P. Govind Raj, "Non Invasive Brain Computer Interface for Controlling User Desktop", available at www.cdacnoida.in/ASCNT-2012/ASCNT-2012/UC/6.pdf, 2012
[12] Hoffmann, Andre, "EEG signal processing & Emotiv's Neuro Head set" Hessen: s. n., 2010
[13] Ruiting Yang, "Signal Processing for Brain Computer Interface", 2009.
[14] Ivan Martinovic, Doug Davies, Mario Frank, Daniele Perito, Tomas Ros, Dawn Song, "On the Feasibility of Side-Channel Attacks with Brain-Computer Interfaces"
[15] Jonathan R. Wolpaw, Niels Birbaumer, Dennis J. McFarland, Gert Pfurtscheller, Theresa M. Vaughan, "Brain Computer Interfaces for Communication and control", Clinical Neurophysiology 113, 767–791, Elsevier, 2002.
[16] Diana Valbuena, Ivan Volosyak, Tatsiana Malechka, Axel Gräsera, "A novel EEG acquisition system for Brain Computer Interfaces", International Journal of Bioelectromagnetism, Vol. 13, No. 2, pp. 74 - 75, 2011
[17] Hoffmann S. and Falkenstein M., "The Corection of eye blinks artifacts in the EEG: A comparison of Two Prominent Methods", PLOS ONE, 2008
[18] Dammess J. Schiek M., Boer F. Silex c., Zvyaginsterv M., Pietrzyk U., and Mathiak K., "Integration of amplitude and phase statstics for complete artifact removal in independent components of neuro magnetic recordings". IEEE Trans Biomed Eng. 55(10):2353-62, 2008
[19] Delmore a., Sejnowski T., and Makeig S., "Enhanced detection of artifacts in EEG data using higher-order statistics & independent component analysis". Neuro image 34(4):1443-9, 2007
[20] Remero S., Mananas M.A., and Barbanoj M.J.."Ocular reduction in EEG signals based on adaptive filtering, regression & blind source separation". Ann Biomed Eng. 37(1):176-91, 2009
[21] Justin Dauwels and Franc¸ois Vialatte, "Topics in Brain Signal Processing", 2010.
[22] Emotiv Systems, available at www.emotiv.com
[23] Neurosky Inc., available at www.neurosky.com
[24] Verleger R.. "The instruction to refrain from blinking affects auditory p3 and N1 amplitudes." Electroencephalogr Chin Neurophysil. 78(3):240-51, 1991
[25] An H D, Po T Wang, Christine E King, Ahmad Abiri and Zoran Nenadic, "Brain-Computer Interface Controlled Functional Electrical Stimulation System for Ankle Movement", Do et al. Journal of Neuro Engineering and Rehabilitation 2011, 8:49., available at http://www.jneuroengrehab.com/content/8/1/49
[26] Government works, available at www.brainwavescience.com/FBI study.php
[27] Tan Le, EPOC Control Panel, available at www.youtube.com/watch?v=G_SRbGXVAGk

**Author's Profile**

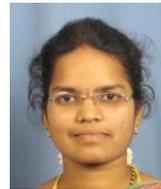
**Ms. M. Rajyalakshmi** received her Masters in Computer Science and Engineering from Jawaharlal Nehru Technological University, Anantapur in 2011. She is currently associated with the Dept. of Comp. Sc. & Engg. at Visvodaya Technical Academy, Kavali, AP, India, as Associate Professor. She has over 5 years of teaching experience at under graduate and graduate level. Her areas of interest are language processing, brain computer interface, Automata, neural networks, etc.

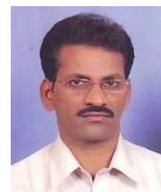
**Mr. T. Kameswara Rao** received his Masters in Computer Applications from University of Madras in 2004 and Masters of Engineering from Satyabhama University in 2007. He is currently associated with the Dept. of Comp. Sc. & Engg. at Visvodaya Technical Academy, Kavali, AP, India, as Associate Professor. He has over 7 years of teaching experience at under graduate and graduate level. His areas of interest are biometrics, brain computer interface, artificial intelligence, neural networks, psychology etc.

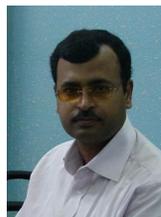
**Dr. T. V. Prasad** has over 17 years of experience in industry and academics. He has received his graduate and master's degree in Computer Science from Nagarjuna University, AP, India. He was with the Bureau of Indian Standards, New Delhi for 11 years as Scientist/Deputy Director. He earned PhD from Jamia Millia Islamia University, New Delhi in the area of computer sciences/ bioinformatics. He has worked as Head of the Department of Computer Science & Engineering, Dean of R&D and Industrial Consultancy and then as Dean of Academic Affairs at Lingaya's University, Faridabad. He is with Visvodaya Technical Academy, Kavali as Dean of Computing Sciences. He has lectured at various international and national forums on subjects related to computing. Prof. Prasad is a member of IEEE, IAENG, Computer Society of India (CSI), and life member of Indian Society of Remote Sensing (ISRS) and APBioNet. His research interests include bioinformatics, consciousness studies, artificial intelligence (natural language processing, swarm intelligence, robotics, BCI, knowledge representation and retrieval). He has over 73 papers in different journals and conferences, and has authored six books and two chapters.